# Comparison of Different Broadcast Schemes for Multi-hop Wireless Sensor Networks[1]

S. Mehta and K.S. Kwak

UWB Wireless Communications Research Center, Inha University
Incheon, 402-751, Korea
suryanand.m@gmail.com

### Abstract

*In this paper, we present the performance of different broadcast schemes for multihop sensor networks based on mathematical modeling. In near future many applications will demand multicast (Broadcast) communication feature from the sensor networks. This broadcast feature does not use virtual carrier sensing but relies on physical carrier sensing to reduce collision. For this paper, we analyze the different broadcast schemes for multihop wireless sensor networks and also calculated the achievable throughput.*

## 1. Introduction

The IEEE 802.11 standard [1] is widely used and deployed in wireless systems. IEEE 802.11 MAC protocol allows multiple nodes to share the wireless medium without any central coordinator. If two nodes that are near by each other transmit frames at the same time, the frames collide and the channel bandwidth will not utilized. The MAC protocol tries to avoid this situation using a mechanism called Multiple Sensing Access with Collision Avoidance (CSMA/CA). CSMA/CA mechanism first listen the channel for a particular duration (a slot time), whenever a node wants to transmit data frame. If the channel is ideal for a particular duration, the node transmits the data frame. The node differ its transmission and waits for a random delay time (back-off interval) before retrying if the channel is busy. This channel sensing mechanism is well-known as physical carrier sensing. Physical carrier sensing does not avoid the collision from the hidden terminal problem if we assume the carrier sensing range is equal to the receiving range. To avoid the collision from hidden terminal, a hidden terminal should not transmit a data frame for particular time period, this time period is known as a vulnerable period. So for broadcasting, only physical carrier sensing is used. Virtual sensing is not directly applicable for broadcast transmission because CTS messages sent by multiple receivers will result in a collision.

All previous studies are for unicast communication; they do not consider broadcast communication. In [2], authors presented mathematical model and a definition for broadcasting scheme, and also the numerical results for IEEE 802.11 DCF. In this paper, we use the same mathematical model and definition of broadcasting as in [2] [5], and also extend the numerical results for slotted aloha and threshold conditions based MAC protocols [ex. IS-MAC] for sensor networks [3][4]. In [4], authors used threshold conditions for transmitting, and these threshold conditions are also useful for broadcast transmission. Here, broadcast transmission refers successful only when all of the sender's neighbors receive the broadcast message correctly.

---

[1] "A part of this paper was published in NEXT 2007 Conference , Seoul,Korea [7]."

Reliable broadcast can be used for number of applications, such as data base application, information distribution, and a basis for supporting distributed protocols. The main contributions of this paper are as follows

- We present the performance of different broadcasting schemes based on mathematical models as in [2][3].
- Our comparison of different broadcasting schemes is very useful for a sensor network designer to set tradeoff between spatial reuse of channel and hidden node area.

The rest of the paper is organized as follows: In section 2, we present the numerical analysis of broadcasting scheme. In section 3, we present numerical results from our analysis. Finally, we conclude in section 4.

## 2. Numerical Analysis

First of all we analyze the performance of IEEE 802.11 broadcast scheme, the hidden node problem in a broadcast scenario, and then we extend our analysis for slotted aloha and threshold conditions based MAC protocols. As shown in figure 1(a), node A is in receiving range of node B but not in the receiving region of node C, may cause hidden terminal problem. This area defined as potential hidden node area. For unicast communications, the size of the potential hidden node area calculated using the distance between sender and receiver. However, same calculation is not applied for broadcast communication. The potential hidden node area for broadcast communication depends on receiving range of all the neighbouring nodes as shown in figure 1(b). So it is difficult to exactly compute the size of this area. Moreover, as explained earlier, varying the carrier sensing area also change the form of this area. When there are infinite numbers of node at the edge of the sender's transmission range, the potential hidden node area is maximized for the worst case. Let $R$ denotes the transmission range of a node. As shown in figure 1(b) maximum size of potential hidden node area can be $\pi(2R)^2 - \pi R^2 = 3\pi R^2$. Thus, in case of broadcast, the potential hidden terminal area can be dramatically larger than that of unicast.

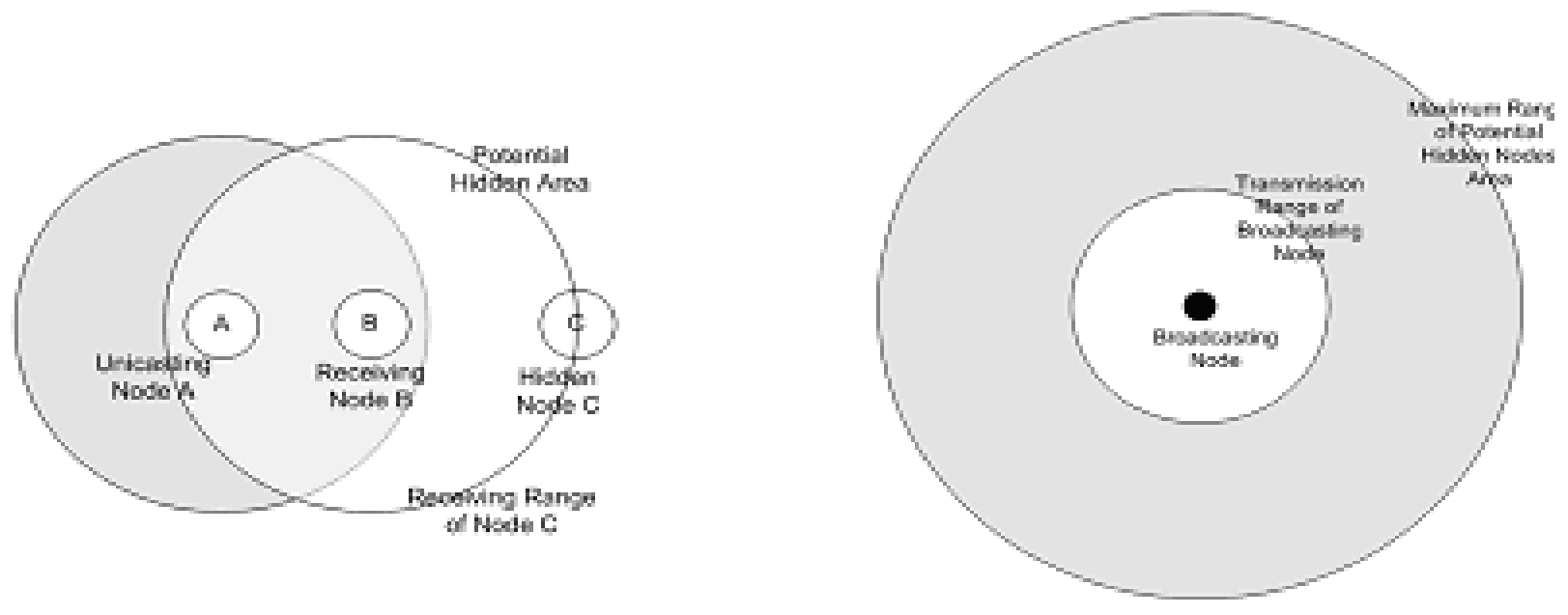

(a) Unicast (b) Broadcast

Figure 1 Potential hidden nodes area

We use the same mathematical models as derived in [2][5] to achieve the average throughput for multihop sensor networks. To make mathematical model tractable, we assume followings for the multi-hop wireless network model.

1. All nodes in the networks are two-dimensionally Poission distributed with density $\lambda$, i.e., the probability $p(i, A)$ of finding $i$ nodes in an area of size $A$ is given by

$$p(i, A) = (\lambda A)^i e^{-\lambda A} / i!$$

2. All nodes have the same uniform transmission and receiving range of radius *R*. *N* is the average number of neighbor nodes within a circular region of $\pi R^2$. Therefore, we have $N = \lambda \pi R^2$.

3. A node transmits a frame only at the beginning of each slot time; however, IS-MAC protocol (Threshold conditions based MAC) based node transmits only if minimum threshold condition gets satisfied [4]. The size of a slot time, $\tau$, is the duration including transmit-to-receive turn-around time, carrier sensing delay and processing time.

4. The transmission time or the frame length is the same for all nodes, i.e., the same data packet length.

5. When a node is transmitting, it cannot receive at the same time.

6. A node is ready to transmit with probability $p$; however, for IS-MAC protocol transmitting probability $p$ also depends on the node's buffered data size [4]. Let $p'$ denote probability that a node transmits in a time slot. If $p'$ is independent at any time slot, it can be obtained by

$$p' = p.\text{Prob}\{\text{channel is sensed idle in a slot}\} \approx p.P_I .$$

Where $P_I$ is the limiting probability that the channel is sensed to be idle.

7. The carrier sensing range is assumed to vary between the range *R~ 2R*.

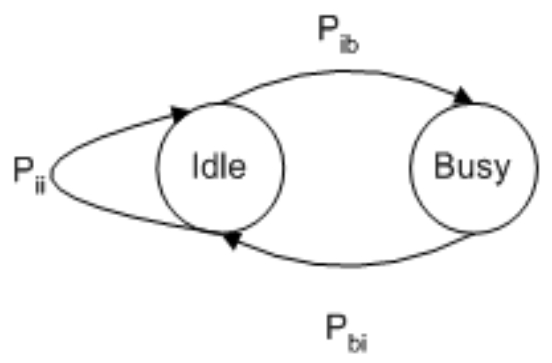

Figure 2 Markov chain model for the channel

From the above mentioned assumptions, the channel process modeled can be represented as a two-state Markov chain shown in figure 2. As shown in the figure 2 this model has 2 states and described as follows

Idle: This is the state when the channel around node '*x*' is sensed idle, and its duration $T_{idle}$ is $\tau$.
Busy: This is the state when a successful DATA transfer is done. The channel is in busy state for the duration of DATA transfer, thus the busy time, $T_{busy}$, is equal to the data transmission time $\delta_{data}$. ($T_{busy} = \delta_{data}$). In MAC scheme, all nodes should stay at least ideal for one slot time, after the channel becomes idle. Thus, the transition probability $P_{bi}$ is 1. $P_{ii}$ is the transition probability of the neighbour nodes transmission, and is given by

$$P_{ii} = \sum_{i=0}^{\infty} \left(1-p'\right)^i \frac{\left(\lambda \pi R^2\right)^i}{i!} e^{-\lambda \pi R^2} = e^{-p'N}$$

Let, $\Phi_i$ and $\Phi_b$ denote the steady-state probabilities of idle and busy states respectively. From figure 2, we have

$$\Phi_i = \Phi_i P_{ii} + \Phi_b P_{bi} = \Phi_i P_{ii} + \Phi_b \ .$$

Since $\Phi_b = 1 - \Phi_i$, we have

$$\Phi_i = \frac{1}{2 - P_{ii}} = \frac{1}{2 - e^{-p'N}}$$

Now the limiting probability $P_I$ can be obtained by

$$P_I = \frac{\tau}{(\delta_{data})(1 - e^{-p'N}) + \tau}$$

According to the relationship between $P'$ and $P$, $P'$ is given by

$$P' = \frac{\tau p}{(\delta_{data})(1 - e^{-p'N}) + \tau}$$

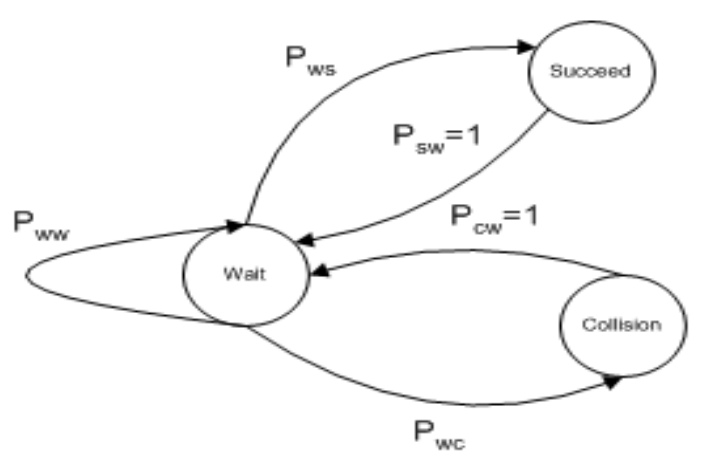

Figure 3 Markov chain model for the node

For the throughput calculation we need to calculate the probability of a successful transmission. As shown in figure 3 the transmission state of node '*x*' can be modelled by three states Markov chain model. Wait, succeed, and collision states represents the node's transmission differ, successful DATA transmission, and collision state conditions, respectively. At the beginning of each time slot, node '*x*' leaves the wait sate with probability $p'$. Thus, the translation probability $P_{ww}$ is given by

$$P_{ww} = 1 - p' \ .$$

And, the duration of a node in wait sate $T_{wait}$ is $\tau$ (This waiting time is only after node satisfy the minimum threshold condition [4]). The duration of success and collision sates are equal to the frame transmission duration time, hence, $T_{succ}$ and $T_{coll}$ are $\delta_{data} + \tau$. After executing the desired action in succeed and collision state, node '*x*' always returns to the wait sate. Therefore, $P_{sw}$ and $P_{cw}$ equals to 1.

Let $\Phi_w, \Phi_s$, and $\Phi_c$ represents the steady-state probabilities of wait, success, and collision states respectively. From the figure 4 we have

$$\Phi_w = \Phi_w P_{ww} + \Phi_s P_{sw} + \Phi_c P_{cw} = \Phi_w P_{ww} + 1 - \Phi_w \; . \qquad (1)$$

Hence, we have

$$\Phi_w = \frac{1}{2 - P_{ww}}$$

Based on the above condition, transition probability $P_{ws}$ cab be

$$P_{ws} = P_1 P_2 P_3 \qquad (2)$$

Where,

$P_1 = \text{Prob}\{\text{node x transmits in a slot}\}$
$P_2 = \text{Prob}\{\text{All of node x's neighbor nodes do not transmit in the same slot}\}$
$P_3 = \text{Prob}\{\text{Nodes in potential hidden area do not transmit for } 2\delta_{data}+\tau\}$

Last term represents the vulnerable period that is equal to $2\delta_{data} + \tau$ and $\delta_{data} + \tau$ in case of slotted aloha based MAC protocols [6][3].Obviously, $P_1 = p'$ and $P_2$ is be given by

$$P_2 = \sum_{i=0}^{\infty} (1-p')^i \frac{(\lambda \pi R^2)^i}{i!} e^{-\lambda \pi R^2} = e^{-p' \lambda \pi R^2} = e^{-p' N}$$

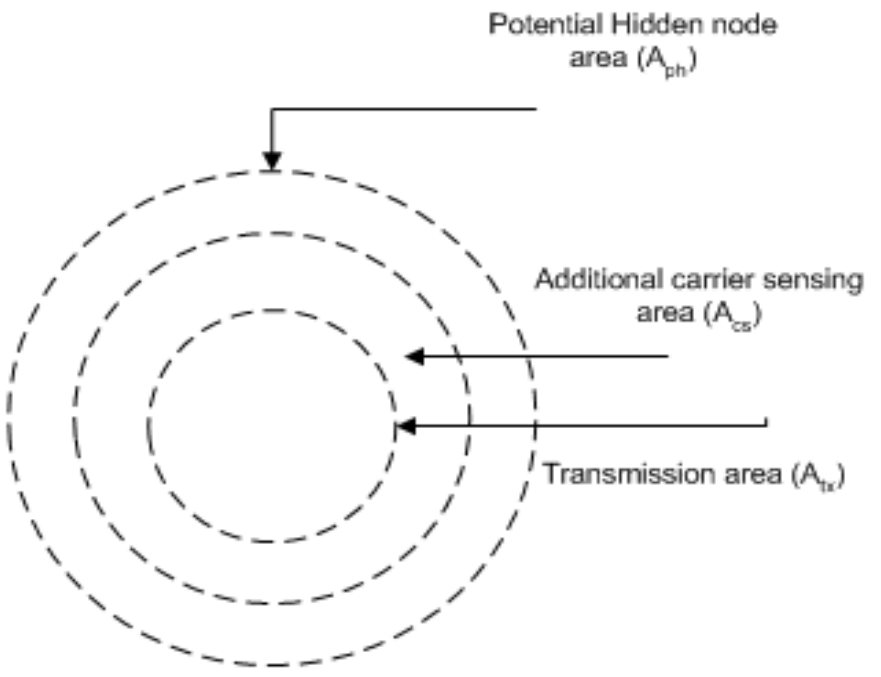

Figure 4 Transmission area, additional carrier sensing area, and potential hidden nodes area

To calculate $P_2$, we first approximate the number of node in the potential hidden node area. Let $A_{tx}$, $A_{cs}$, and $A_{ph}$ represent the transmission area, additional carrier sensing area, and potential hidden node area, respectively. As shown in figure 4 additional carriers sensing area is the physical carrier sensing area is larger than transmission range and smaller than potential hidden node area. Hence, we have

$$0 \le A_{cs} \le 3\pi R^2 \; .$$

And, the potential hidden node area is given by

$$A_{ph} = 2\pi\left(2R\right)^2 - A_{tx} - A_{cs} = 2\pi(2R)^2 - \pi R^2 - A_{cs} = 3\pi R^2 - A_{cs} .$$

Hence,

$$0 \le A_{ph} \le 3\pi R^2 .$$

Let $N_{ph}$ represent the number of node in potential node area. As we assumed, nodes are uniformly distributed, hence, $N_{ph}$ is given by

$$\begin{aligned} N_{ph} &= \lambda A_{ph} \\ 0 \le N_{ph} &\le \lambda 3\pi R^2 \\ 0 \le N_{ph} &\le 3N \end{aligned} \qquad (3)$$

With eq.3 $P_3$ is given by

$$P_3 = \left\{ \sum_{i=0}^{\infty} (1-p)^i \frac{(N_{ph})^i}{i!} e^{-N_{ph}} \right\}^{(2\delta_{data}+\tau)} = e^{-p' N_{ph}(2\delta_{data}+\tau)}$$

Therefore, eq.2 is given by

$$P_{ws} = p' e^{-p'N} e^{-p'3N_{ph}(2\delta_{data}+\tau)} .$$

From the figure 4, we have $P_{ws} = 1 - P_{ww} - P_{wc}$ and $P_{cw} = P_{sw} = 1$. Hence, the steady state probability of succeed state, $\Phi_s$, is given by

$$\Phi_s = \Phi_w P_{ws} = \frac{P_{ws}}{1+p'}$$

According to the definition of throughput from [6] , the throughput equals the fraction of time in which the channel is engaged in successful transmission of user data. Therefore, the throughput $Th$ is equal to the limiting probability that the channel is in success state.

$$Th = \frac{\Phi_s \delta_{data}}{\Phi_s T_{succ} + \Phi_c T_{coll} + \Phi_w T_{wait}} \qquad (4)$$

$$Th = \frac{P_{ws}\delta_{data}}{p' T_{coll} + T_{wait}} = \frac{(p' e^{-p'(N+3N_{ph}(2\delta_{data}+\tau))})\delta_{data}}{\tau + p'(\delta_{data}+\tau)}$$

## 3. Numerical Results

In this section, we present numerical results based on the model presented in the previous section. We first study the performance of IEEE 802.11 broadcast scheme by varying the average number of neighbouring node (*N)* and transmission attempt probability (*p'*). In this analysis, we fix the data frame as $100\tau$ . Figure 5 shows the throughput results of the IEEE 802.11 broadcast scheme with different potential hidden node area ( $R_{ph}$ ). From the figure 6 it is cleared that, as the percentage of $A_{ph}$ reduces to 0, throughput performance increases to maximum value. This means that, by achieving maximum value of $A_{cs}$, the IEEE 802.11 broadcast scheme minimizes the probability of hidden node problem. Figure 6 shows the throughput results of the threshold conditions based broadcast scheme with different potential hidden node area. These threshold conditions help to improve the throughput by not letting all the nodes to transmit in the same slot at the same time. With these threshold conditions, we achieve nearly twice of the throughput compared to IEEE 802.11 broadcast scheme. Figure 7 shows the throughput results of the slotted aloha based broadcast scheme. Slotted aloha based broadcast scheme with threshold conditions give quite good throughput. Figure 8 shows the combined results of all the schemes with all the variations in hidden node area. From figure 9 it is clear that, the threshold conditions based broadcast scheme with 0% hidden area gives the highest throughput. So it is beneficial to set the large carrier sensing range for broadcast communication. However, for unicast communication, a large carrier sensing range leads to reduce spatial reuse, so minimizing hidden node effect and increasing spatial reuse becomes a tradeoff which must be studied further. Slotted aloha and threshold conditions based broadcast schemes help us to achieve a good tradeoff between spatial reuse and hidden node effect. Our results reveal the performance of broadcasting communications under the impact of hidden terminals and open some new directions for further research.

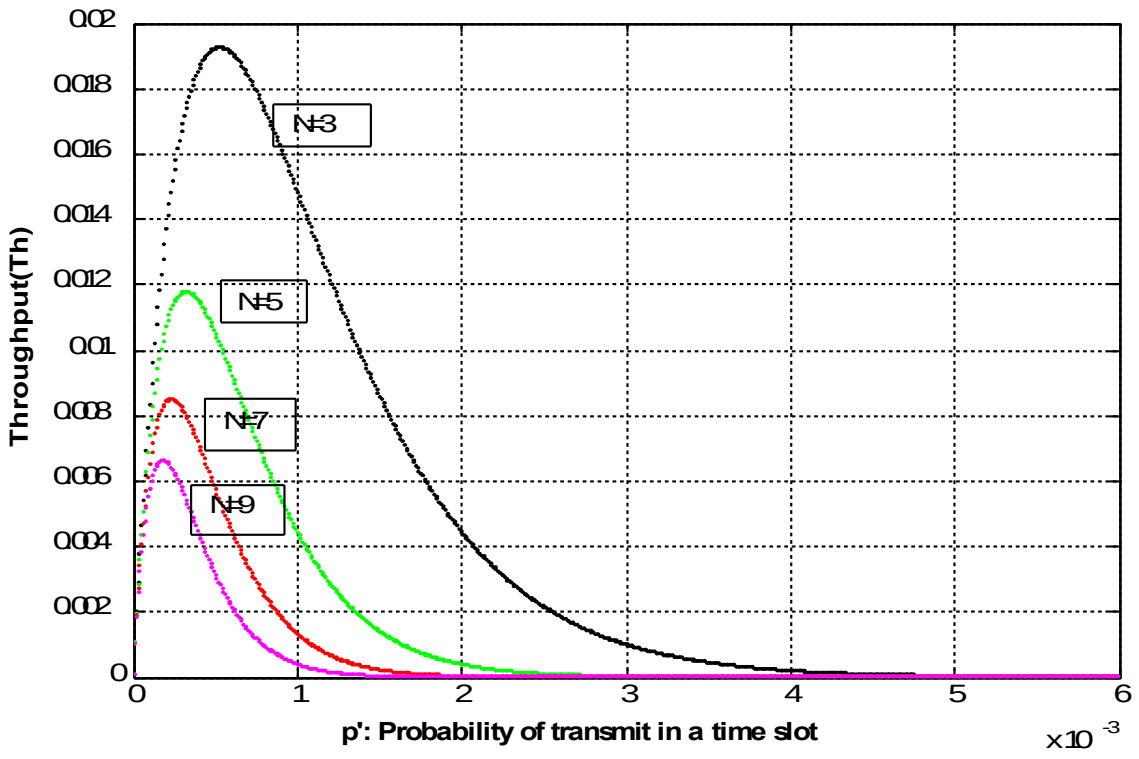

(a) 100% Hidden node area

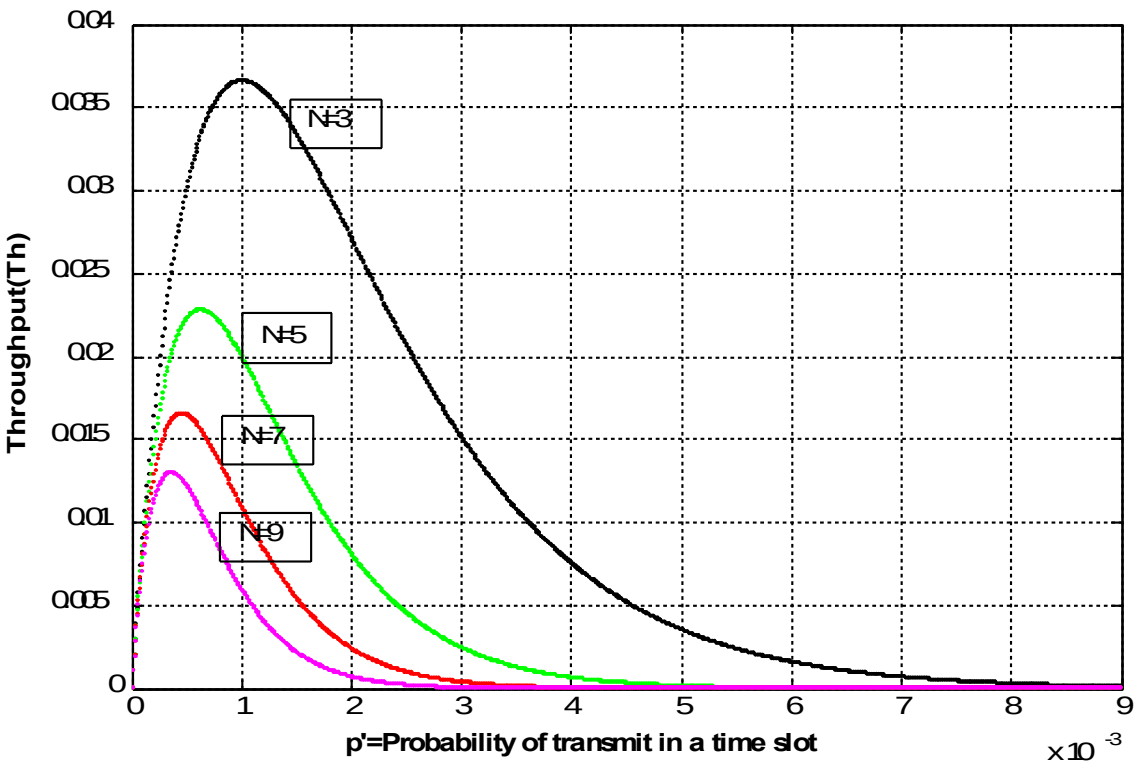

(b) 50% Hidden node area

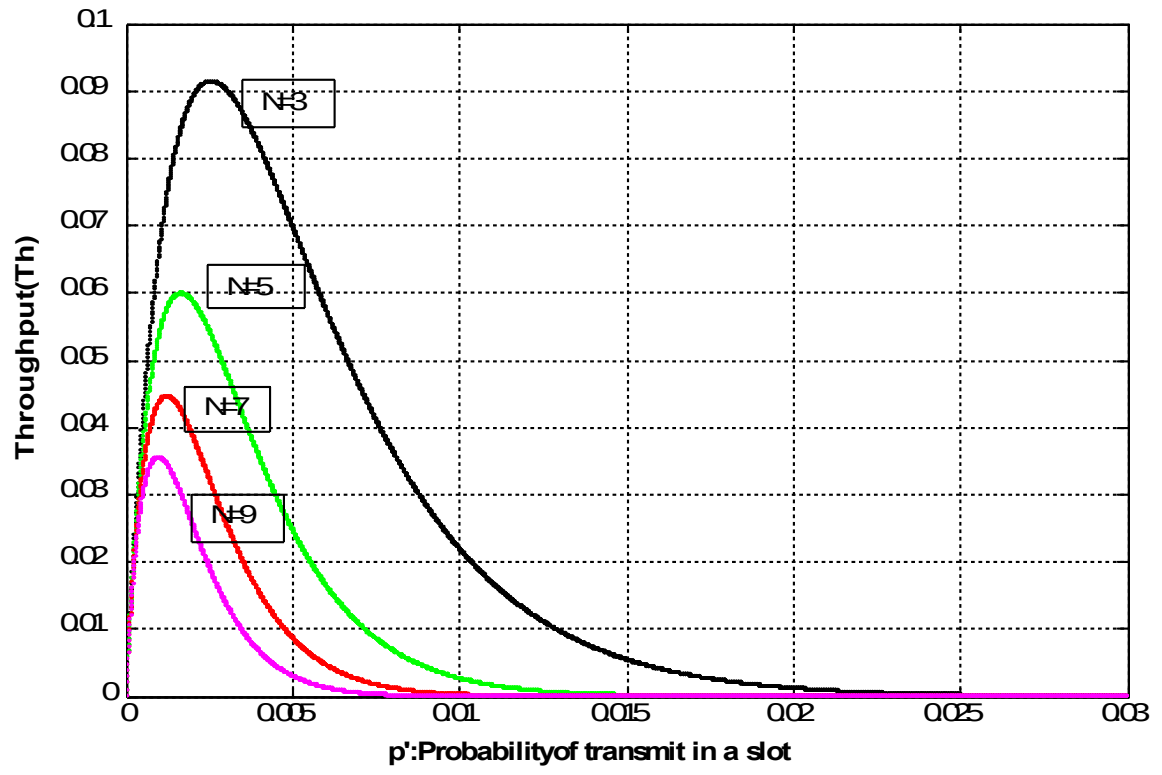

(c) 0% Hidden node area

Figure 5 IEEE 802.11 based broadcast scheme

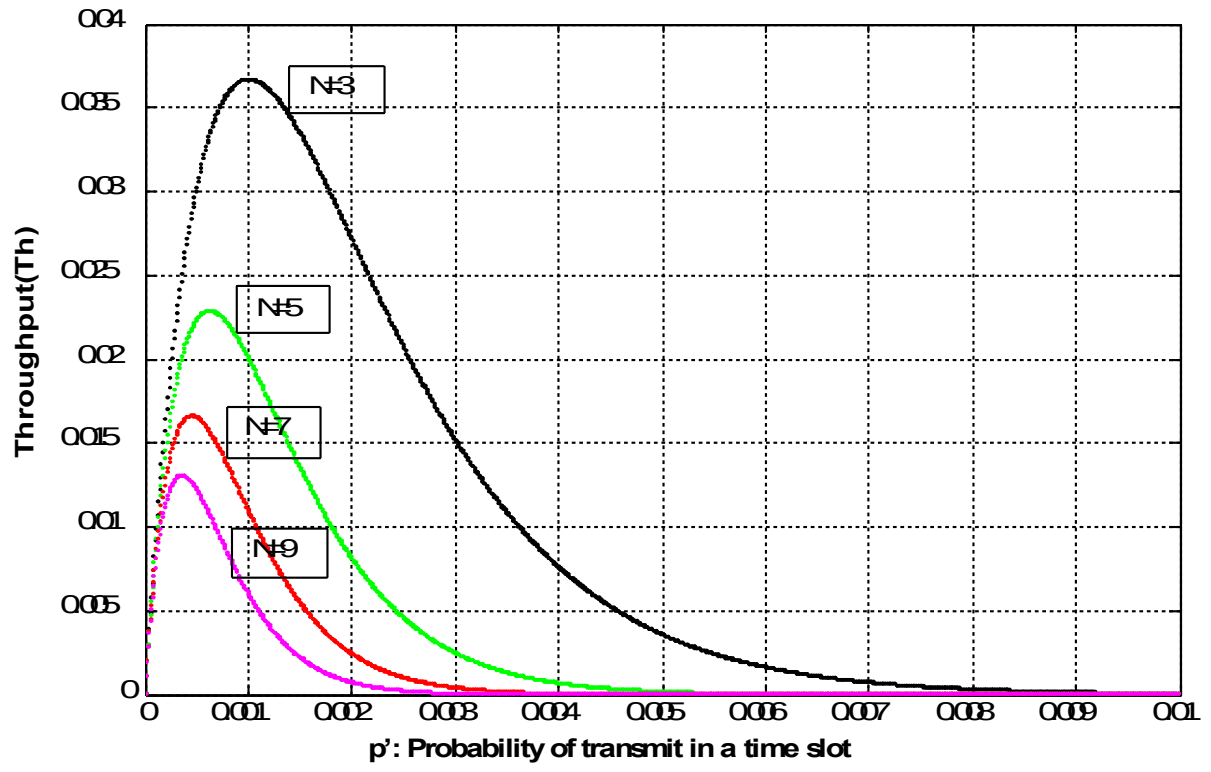

(a) 100% Hidden node area

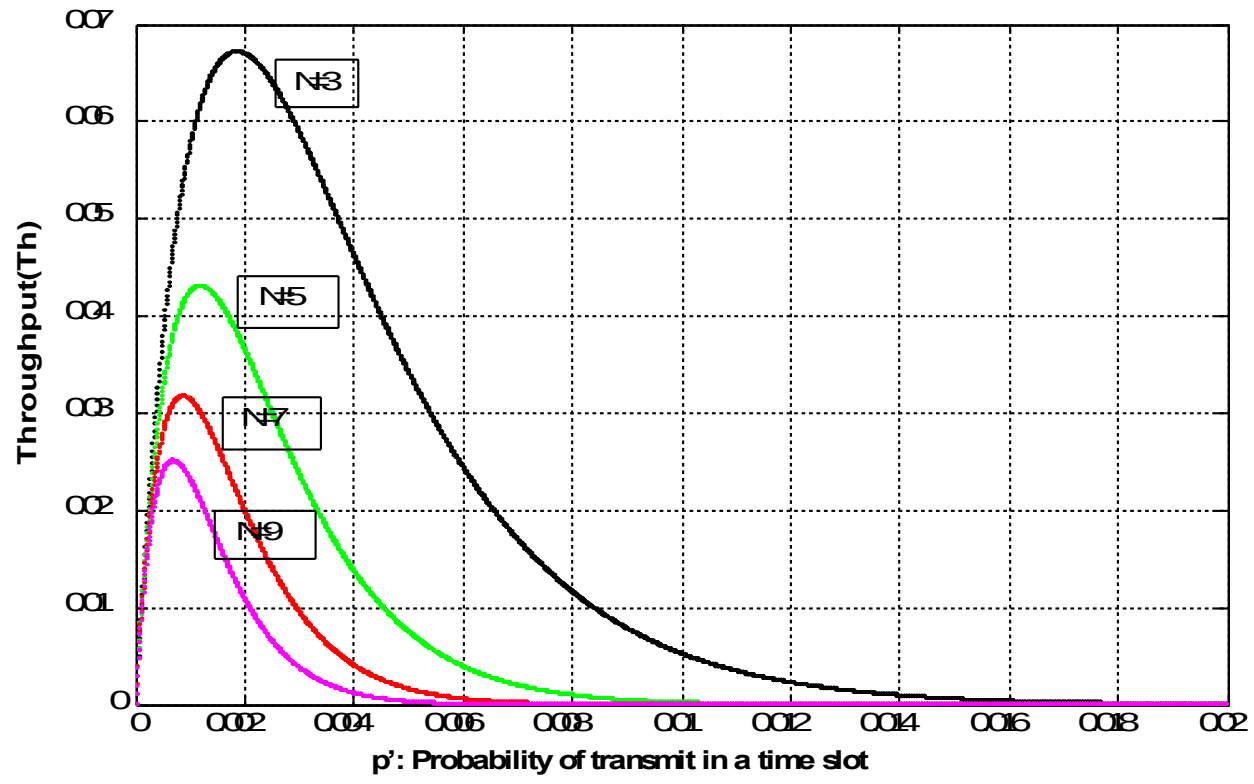

(b) 50% Hidden node area

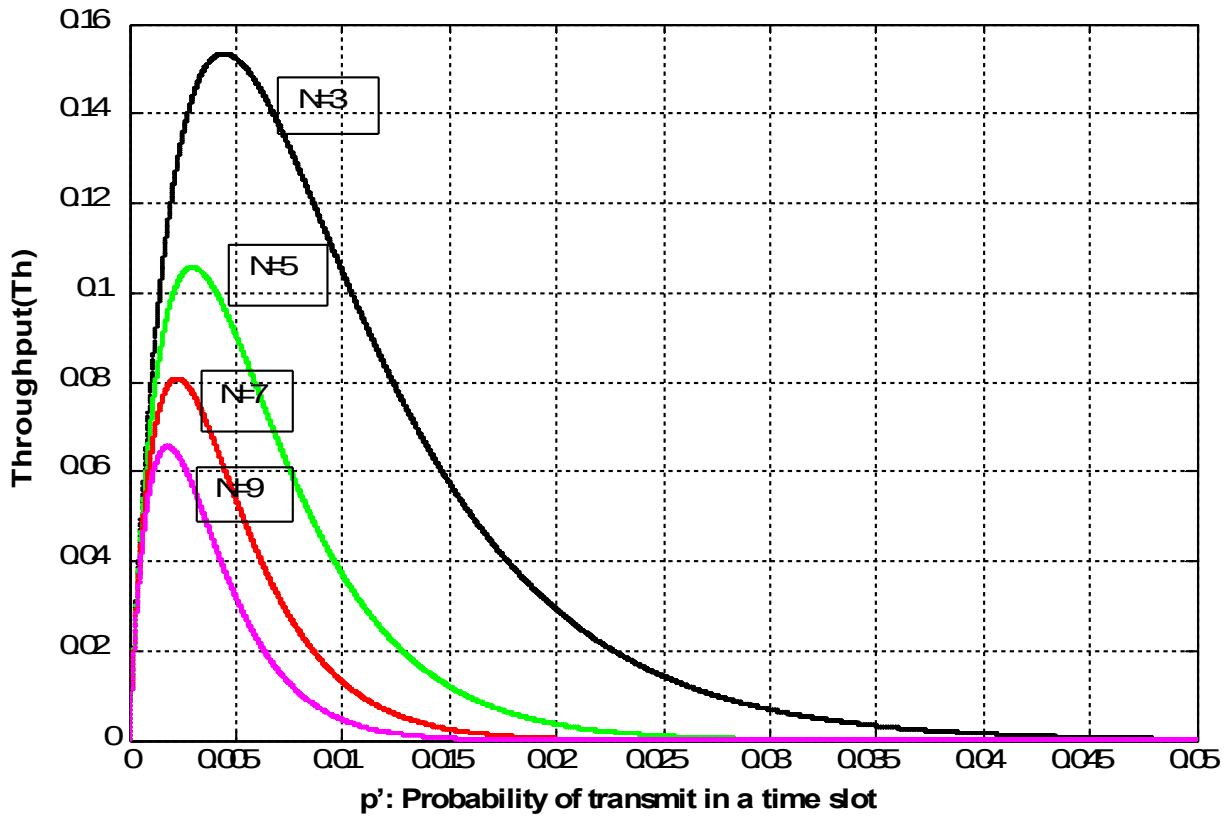

(c) 0% Hidden node area

Figure 6 Threshold conditions based broadcast scheme

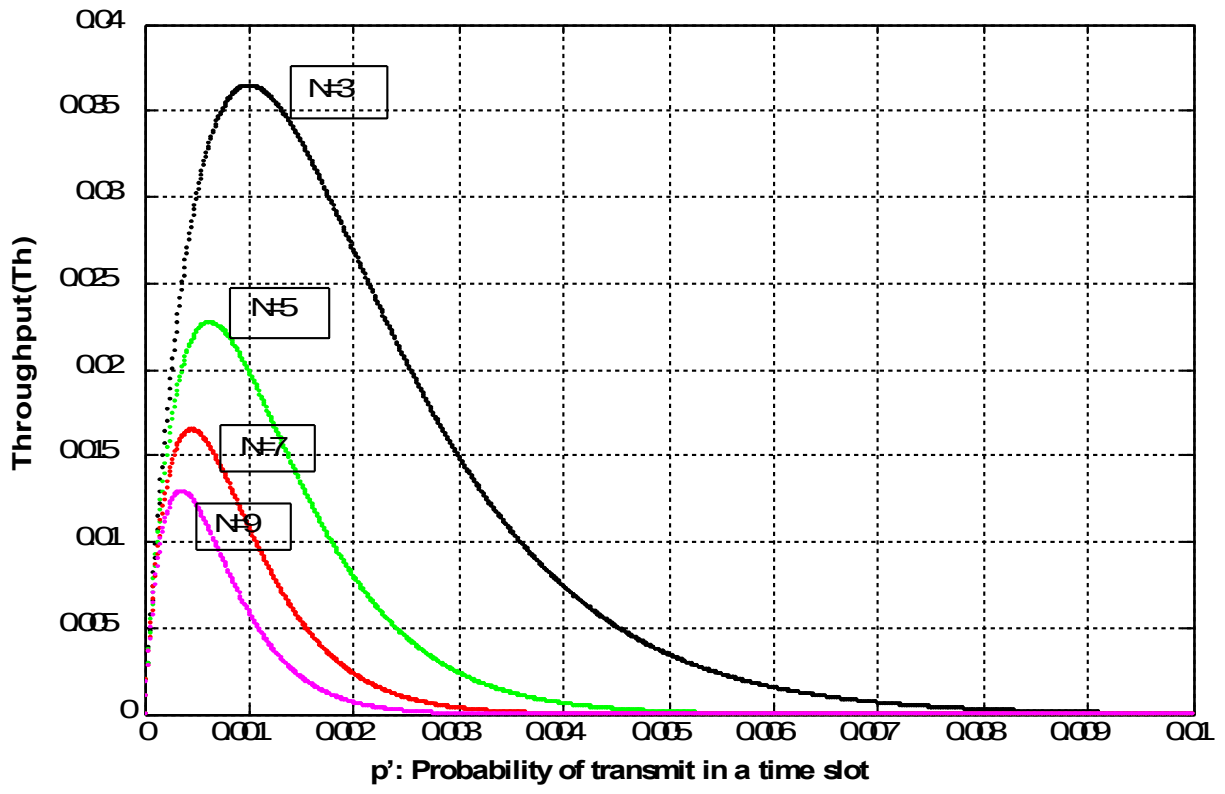

(a) Slotted aloha type scheme without threshold conditions

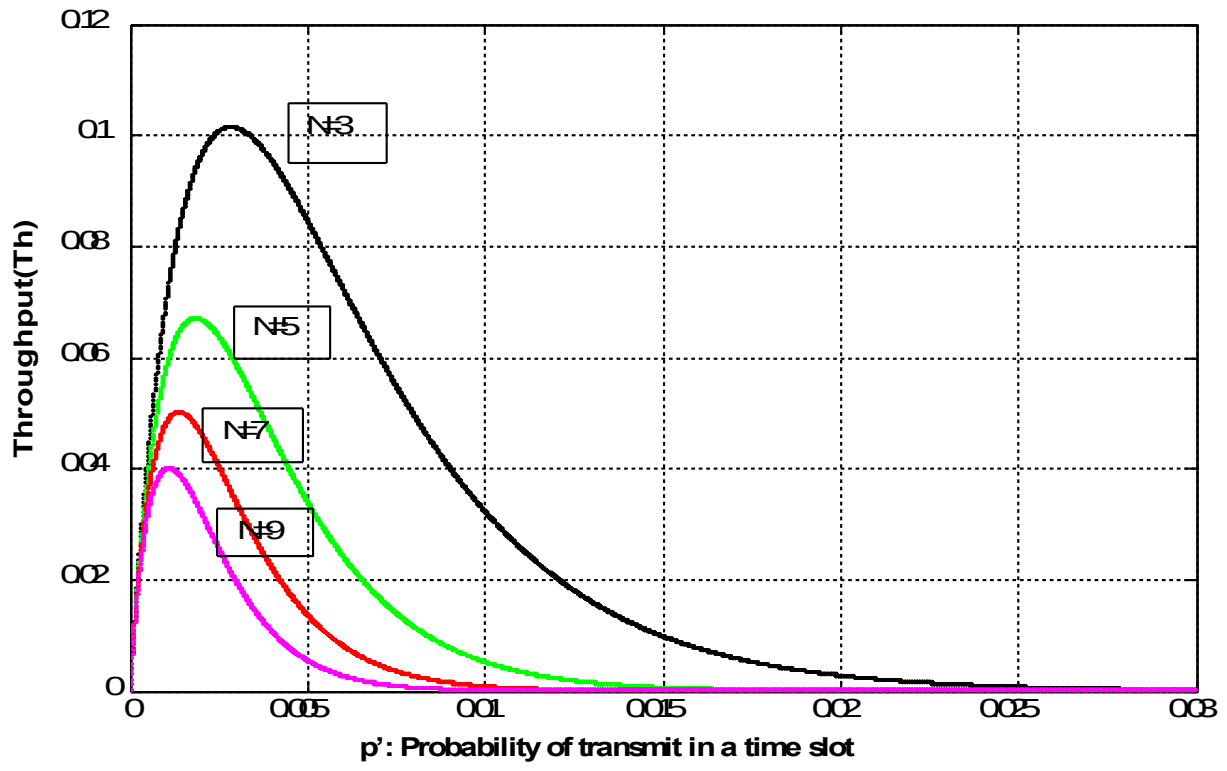

(b) Slotted aloha type scheme with threshold conditions

Figure 7 Slotted aloha type broadcasting scheme

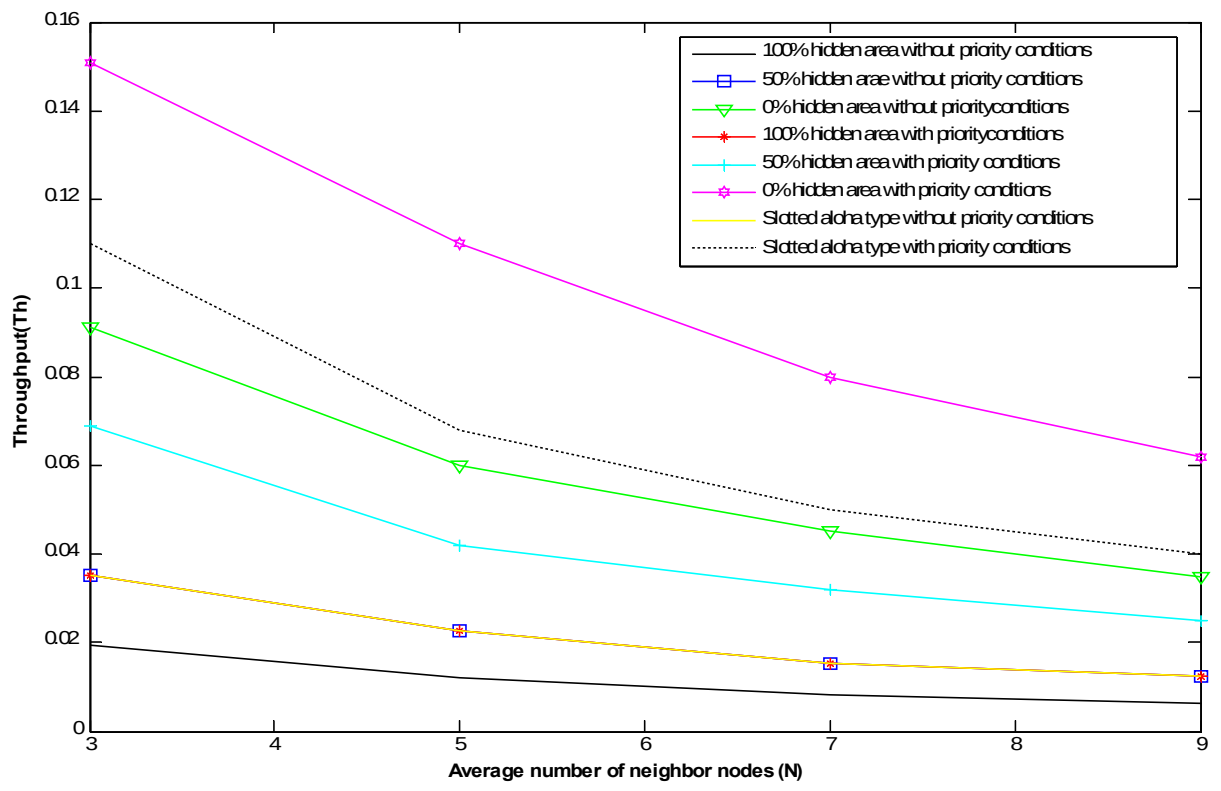

Figure 8 Combined results

## 4. Conclusions

In this paper, we present the performance of different broadcasting schemes based on a simple markov chain model. The results show that overall performance of different broadcasting schemes degrades rapidly when the number of competing nodes allowed within a region increases. Our comparison of different broadcasting schemes is very useful for sensor network designer, and also helps us to set tradeoff between spatial reuse of channel and hidden node area. In future, we want to extend our study for multi-channel hidden terminals and non-uniform receiving/transmitting range problems of a broadcasting node.

## Acknowledgement

This research was supported by the MKE(The Ministry of Knowledge Economy), Korea, under the ITRC(Information Technology Research Center) support program supervised by the NIPA(National IT Industry Promotion Agency) (NIPA-2010-C1090-1011-0007).

**Authors**

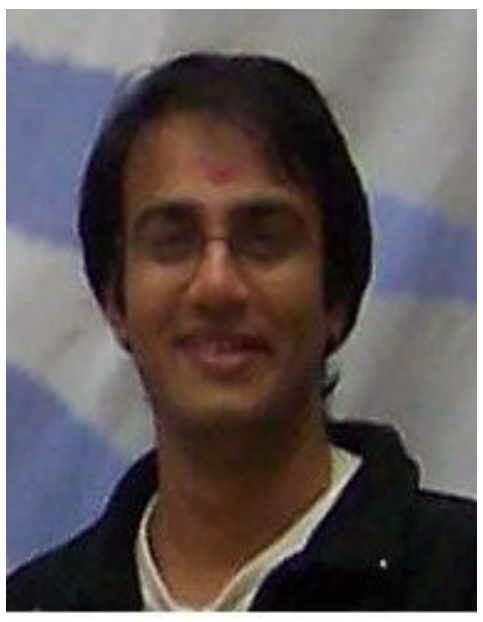

**S.Mehta** received the B.E. and M.S degrees both in Electronics Engineering from Mumbai University, Mumbai, India, and Ajou University, Korea in 2002 and 2005, respectively. He is currently pursuing the Ph.D. degree in Telecommunication engineering from the Inha University, Korea. His research interests are in the area of performance analysis of wireless networks and RFID systems.

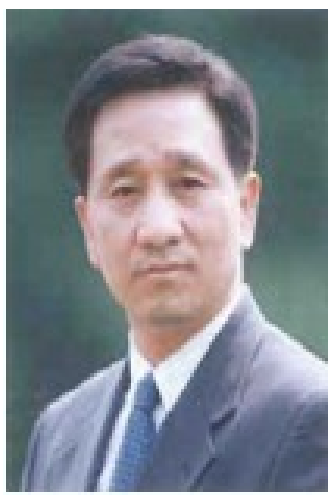

**K. S. Kwak** received the B.S. degree from Inha University, Korea in 1977, and the M.S. degree from the University of Southern California in 1981 and the Ph.D. degree from the University of California at San Diego in 1988, respectively. From 1988 to 1989 he was a Member of Technical Staff at Hughes Network Systems, San Diego, California. From 1989 to 1990 he was with the IBM Network Analysis Center at Research Triangle Park, North Carolina. Since then he has been with the School of Information and Communication, Inha University, Korea as a professor. He had been the chairman of the School of Electrical and Computer Engineering from 1999 to 2000 and the dean of the Graduate School of Information Technology and Telecommunications from 2001 to 2002 at Inha University, Inchon, Korea. He is the current director of Advanced IT Research Center of Inha University, and UWB Wireless Communications Research Center, a key government IT research center, Korea. He has been the Korean Institute of Communication Sciences (KICS)'s president of 2006 year term. In 1993, he received Engineering College Young Investigator Achievement Award from Inha University, and a distinguished service medal from the Institute of Electronics Engineers of Korea (IEEK). In 1996 and 1999, he received distinguished service medals from the KICS. He received the Inha University Engineering Paper Award and the LG Paper Award in 1998, and Motorola Paper Award in 2000. His research interests include multiple access communication systems, mobile communication systems, UWB radio systems and ad-hoc networks, high-performance wireless Internet.